\documentstyle[12pt]{article}
\parskip=\medskipamount                 
\thicklines                         

\begin{document}

\font\blackboard=msbm10 scaled 1200 \font\blackboards=msbm7 scaled 1200
\font\blackboardss=msbm5 scaled 1200
\font\fblac=msbm10 scaled 1000
\newfam\black
\textfont\black=\blackboard
\scriptfont\black=\blackboards
\scriptscriptfont\black=\blackboardss
\def\blackb#1{{\fam\black\relax#1}}
\newfam\fblack
\textfont\fblack=\blackboard
\scriptfont\fblack=\fblac
\scriptscriptfont\fblack=\blackboardss
\def\fbb#1{{\fam\fblack\relax#1}}
%

%
\def\BC{{\blackb C}} \def\BQ{{\blackb Q}}
\def\BR{{\blackb R}} \def\BI{{\blackb I}}
\def\BZ{{\blackb Z}} \def\BD{{\blackb D}}
\def\BP{{\blackb P}}
\def\BS{{\blackb S}}
\renewcommand{\thefootnote}{\fnsymbol{footnote}}
\begin{flushright}
IMSc--94/28~~~~\\
TIFR/TH/94-14\\
hep-th/9405146\\
Revised October 1994
\end{flushright}
\bigskip
{\centerline{\large \bf A proposal for the geometry of $W_n$ gravity}}
\bigskip\bigskip
\centerline{Suresh Govindarajan\footnote{E-mail: suresh@theory.tifr.res.in}}
\centerline{Theoretical Physics Group}
\centerline{Tata Institute of Fundamental Research}
\centerline{Bombay 400 005 INDIA}
\bigskip
\centerline{T. Jayaraman\footnote{E-mail: jayaram@imsc.ernet.in}}
\centerline{The Institute of Mathematical Sciences}
\centerline{C.I.T. Campus, Taramani}
\centerline{Madras 600 113, INDIA}
\bigskip 
\noindent {\bf Abstract:} We relate the Teichmuller spaces obtained by Hitchin
to the Teichmuller spaces of $WA_{n}$-gravity. The relationship of this space
to $W$-gravity is obtained by identifying the flat $PSL(n+1,{\BR})$
connections of Hitchin to gen\-eralis\-ed viel\-beins and
conn\-ec\-tions. This is explicitly demonstrated for $WA_2=W_3$
gravity. We show how $W$-diffeomorphisms are obtained in this
formulation. We find that particular combinations of the generalised
connection play the role of projective connections. We thus obtain
$W$-diffeomorphisms in a geometric fashion without invoking the
presence of matter fields.  This description in terms of vielbeins
naturally provides the measure for the gravity sector in the Polyakov
path integral for $W$-strings.
\newpage
\renewcommand{\thefootnote}{\arabic{footnote}}
\setcounter{footnote}{0}
\section{Introduction}

One of the ways new structure arises in string theory is by extending
the gauge symmetry of the world sheet beyond bosonic reparametrization
invariance. The most well-known is of course supersymmetry, where
there are both bosonic and fermionic degrees of freedom on the world
sheet leading to local supersymmetry on the world sheet. The addition
of this new symmetry plays an important role in non-critical strings
allowing the extension of string theories beyond the bosonic
strong-coupling barrier of $c=1$. It is interesting to examine whether
there are new structures associated with bosonic and/or fermionic
chiral algebras (in the CFT sense) apart from the Virasoro algebra.
This has been one of the prime motivations in studying $W$-strings
where the $W$-algebra is the additional chiral algebra present in the
theory.

Among the outstanding problems of formulating a theory of $W$-strings
are first the non-linearity of the algebra\cite{zam} and the lack of
knowledge of the structure of the associated moduli and Teichmuller
spaces, especially the measure.  In the Polyakov path integral
formulation of string theory, one can show that the integration over
all metrics can be reduced to one over the moduli space of Riemann
surfaces.  Associated to this moduli space is the Teichmuller space
which when quotiented by the mapping class group gives moduli
space. Not much is known about this space in the case of $W$-strings
except for the dimension, obtained by counting the number of zero
modes of the associated ghost systems.  The other problem is of course
the insufficient understanding of $W$-coordinates which is especially
difficult given the bosonic nature of the $W$-symmetry.  Several
attempts have been made in the literature to tackle these issues but
the problems remain largely unresolved. In this paper we make an
attempt to tackle the first problem, basing ourselves on the work of
Hitchin on Higgs bundles\cite{hitchin}.

In the case of the bosonic string, it is known that the Teichmuller
space can be obtained as a particular component of the space (for a
surface $\Sigma$ with genus $g>1$) ${\cal T}_2(\Sigma)\equiv$
Hom$(\pi_1(\Sigma)\rightarrow
PSL(2,\BR))/PSL(2,\BR)$. Hitchin\cite{hitchin} has studied the spaces
where $PSL(2,\BR)$ is replaced by an arbitrary semi-simple group
$G$. We will be interested in the case when $G=PSL(n,\BR)$. This space
has many disconnected components which includes a certain component
(called the Teichmuller component) which has dimension
$(2g-2)~dim(G)$. For $G=PSL(n,\BR)$, we denote the Teichmuller
component by ${\cal T}_n(\Sigma)$. We would first like to observe that
the dimension of the Teichmuller space of $WA_{n-1}$-strings is the
same as that of ${\cal T}_n(\Sigma)$.  Second, for the case $n=2$,
this reduces to the usual Teichmuller space as we have already
observed.  We propose that the Teichmuller spaces ${\cal T}_n(\Sigma)$
introduced by Hitchin are the Teichmuller spaces for $W$-gravity.

Hitchin obtains the Teichmuller spaces as the space of solutions of
the {\it self-duality} equations.  These equations can be obtained
from four dimensional self-dual Yang-Mills by dimensional
reduction\cite{selfdual}. The self-duality equations are
\begin{equation}
\label{eselfdual}
F_A + [\Phi , \Phi^\dagger] = 0 \quad,
\end{equation}
where $A$ is a unitary connection on a holomorphic vector bundle $V$,
$F_A$ its field strength and $\Phi$ is a holomorphic section of
End$~V\otimes K$. $\Phi$ is called a Higgs field
and $K$ is the canonical line bundle. The holomorphicity condition on
$\Phi$ is
\begin{equation}
d_A''~\Phi=0\quad. \label{ehol}
\end{equation}
It is important to note that the self-duality equations are
conformally invariant, i.e., they only depend on the conformal class
of the metric.  The different Teichmuller spaces obtained by Hitchin
correspond to various choices of the Higgs pair $(V,\Phi)$.
For example, in the $PSL(2)$ case, $V=(K^{-{1\over2}}\oplus
K^{1\over2})$ and $\Phi=\pmatrix{0&h~dz\cr0&0}$. In this case,
the self-duality equations (\ref{eselfdual}) can be shown to be
equivalent to the constant negative curvature condition\cite{selfdual}.

A first attempt to relate Higgs bundles to W-gravity was made by de
Boer and Goeree (dBG)\cite{dbg}.  dBG pointed out that the Higgs
bundle setting was the way to generalise their ``covariant"
constructions for pure $W_n$-gravity to higher genus Riemann surfaces.
However, their construction is somewhat indirect and involves the
introduction of a somewhat arbitrary ``reference connection''.

In this letter, we attempt to establish that the spaces ${\cal T}_n(\Sigma)$
are
indeed the Teichmuller spaces of $W_n$ gravity. We begin directly in the Higgs
bundle setting, by identifying
components of the Higgs field with a generalisation of the vielbein
and the $SO(n)$ connection with a generalisation of the spin -
connection. Conceptually this is an important point since it provides
a way of understanding various formulations of $W_n$-gravity since the
vielbeins and spin-connections give a covariant description of
$W_n$-gravity.  In principle, this gives a means of obtaining the
equivalent of ``general" coordinate transformations for W-gravity. We
can also obtain a direct relationship to the ``metric'' formulations
of W-gravity a la Hull and others\cite{hull}. This is in a sense a
dual picture of the construction of W-generators as higher order
Casimirs\cite{bouwknegt}.  Projective connections are obtained as
geometric constructs without any direct reference to matter. This is
essential for a geometric picture.  Finally, we suggest that just as
the uniformisation of
Riemann surfaces is related to flat $PSL(2,\BR)$ connections, there must exist
such a picture for the $W_n$-case.

In another paper\cite{falqui}, Aldrovandi and Falqui have discussed
the relation of Higgs bundles to Toda systems and hence through that
to formulations of W-gravity in a ``liouville"-like formulation. They
have also recognised the relationship of Higgs bundles to an
uniformisation problem.  However we believe that in our formulation of
the problem there is a more explicit geometric structure that is
revealed, that is also more general.

The paper is organised as follows. In section 2, we use the bosonic
string to illustrate how the identification of vielbein with the Higgs
field and the connection $A$ with the spin connection works.
In section 3, we discuss
the $W_n$ case using the $W_3$ as an explicit example. In section 4,
we relate to the works of \cite{GLM,BFK,others,dbg} by explicitly
obtaining $W$-diffeomorphisms in our formulation. Finally, in section
5, we conclude with a brief discussion on these results and provide
some suggestions.
\section{The bosonic string}
In this section, we shall demonstrate how our identifications of the
vielbein as well as the spin-connection works in the case of the
bosonic string. In addition, this will also illustrate self-duality
equations. Consider, the case where the vector bundle
$V=K^{-{1\over2}}\oplus K^{{1\over2}}$ and
\begin{equation}\label{ehiggsa}
\Phi=\pmatrix{0&h~dz\cr0&0}\quad.
\end{equation}
  Note that $dim(V)=2$ which implies that the connection $A$ in the
self-duality equations is a $SU(2)$ connection\footnote{The existence
of such a connection requires that a certain stability condition has
to be satisfied\cite{selfdual}. For all the situations we discuss in
this paper, we have a stable Higgs bundle\cite{hitchin}.}.
Substituting for $\Phi$ in the self-duality equations gives
\begin{equation}\label{esolve}
F_A = \pmatrix{{1\over2}F^0 & 0 \cr 0 &{-{1\over2}}F^0}
    =\pmatrix{{-1} & 0 \cr 0 & 1} h^2 dz\wedge d{\bar z}\quad.
\end{equation}
One can easily see from (\ref{esolve}) that since $F^\pm=0$. This
implies that the $SU(2)$ connection is reducible to a $U(1)$
connection. This $U(1)$ connection can be identified with the
spin-connection. It then follows that $F^0$ corresponds to the
curvature and hence (\ref{esolve}) is the constant negative curvature
condition.

The connection ${\cal A}$ defined by $$ {\cal
A}\equiv(A+\Phi+\Phi^\dagger) $$ is a flat $SL(2,\BC)$ connection. The
flatness condition follows from the self-duality equation
(\ref{eselfdual}) and the condition that $\Phi$ is holomorphic
(Eqn. (\ref{ehol})).  Corlette and Donaldson have proved the converse
of the above statement\cite{corlette,donald}. They have shown that if
$V$ is a vector bundle on a Riemann surface $\Sigma$ with a completely
reducible flat connection ${\cal A}$, then there exists a section
(gauge), where ${\cal A} = (A + \Phi +
\Phi^\dagger)$ where $(A,\Phi)$ satisfy the self-duality equations.

Given a flat connection with gauge group $G$, it can be used to
construct a map from $\pi_1(\Sigma)$ to the group $G$ by using the
Wilson loop operator. The flatness of the connection ensures that the
map depends only on the homotopy class of the path. See
\cite{tftreview} for a discussion. Since the Wilson loop is invariant
under conjugation, we obtain an element of
Hom$(\pi_1(\Sigma)\rightarrow G)/G$.

Returning to our example, Hitchin has shown that for this choice of
Higgs field, the holonomy data is contained in real form
(isomorphic to $PSL(2,\BR)$ ) of $PSL(2,\BC)$. Hence, for
purposes of the holonomy data, we can represent the connection ${\cal
A}$ by a flat $PSL(2,\BR)$ connection. In summary, the Higgs bundle
is associated to a flat $PSL(2,\BR)$ connection.

Now modify the Higgs field to
\begin{equation}\label{ehiggsb}
\Phi=\pmatrix{0&h\cr {a\over h}&0}\quad,
\end{equation}
where $a\in$Hom$(K^{-{1\over2}},K^{1\over2})\otimes K
\sim K^2$. Hence, $a$ is a holomorphic quadratic differential. Again,
the self-duality equations give the constant negative curvature
condition\cite{selfdual}. We know that the space of constant negative
curvature metrics is the Teichmuller space. We have also shown that
the space of self-duality equations with $\Phi=\pmatrix{0&h\cr {a\over
h}&0}$
gives the space of constant curvature metrics on $\Sigma$ and hence is
gives Teichmuller space.

We have already identified the $U(1)$ connection with the
spin-connection. We now identify the Higgs field with the vielbein
of the Riemann surface. This identification was first made
in \cite{tftreview}. The $PSL(2,\BR)$ connection can then be explicitly
written in terms of the vielbein and spin-connection as
\begin{equation} \label{eident}
{\cal A} =\pmatrix{{\omega\over2}&{e^+\over\sqrt{2}}\cr{e^-\over\sqrt2}&
{-{\omega}\over{2}}}\quad,
\end{equation}
where $\omega$ is the spin-connection and $e^\pm$ are the
vielbein. The flatness condition on the vielbein are nothing but the
usual torsion constraints imposed on them. Comparing with the explicit
form of $\Phi$ in eqn. (\ref{ehiggsa}), we see that the flatness
condition reduces to the self-duality equations in the ``conformal
gauge'' where $e^+_z=e^-_{\bar z}=h$, $e^-_z = e^+_{\bar
z}=0$. $(+,-)$ are the tangent space indices and $(z,~{\bar z})$ are
Einstein indices. The fact that the vielbein and the spin-connection
can be combined into a $PSL(2,\BR)$ gauge field (provided the constant
negative curvature condition as well as the torsion constraints are
imposed) has been observed in \cite{topgrav,verlinde}.  The
$PSL(2,\BR)$ transformations given below
\begin{eqnarray}
\delta~e^+_\mu = \partial_\mu\epsilon^+-\omega_\mu\epsilon^++\alpha
e^+_\mu \nonumber \quad,\\
\delta~e^-_\mu = \partial_\mu\epsilon^-+\omega_\mu\epsilon^--\alpha e^-_\mu
\nonumber \quad,\\
\delta~\omega_\mu = \partial_\mu \alpha - e^+_\mu\epsilon^-
+ e^-_\mu\epsilon^+ \quad,
\end{eqnarray}
then correspond to diffeomorphisms (with parameters $\xi^\mu$)
provided we choose
$$
\epsilon^a=\xi^\mu e_\mu^a\quad,\quad \alpha=\xi^\mu\omega_\mu\quad.
$$
The flatness conditions are now given below. As can be clearly seen,
the first expression are the usual torsion free condition on the
vielbein.
\begin{eqnarray}\label{eflata}
T^\pm &\equiv& d~e^+ \mp \omega\wedge e^+ =0\quad,\nonumber \\
F^0 &\equiv& d~\omega =- e^+ \wedge e^- \quad.
\end{eqnarray}
In the vielbein formulation, the metric is given by
\begin{equation}\label{emetric}
g_{\mu\nu} = e_\mu^a~\delta_{ab}~e_\nu^b \sim tr(E^2)\quad,
\end{equation}
where $E\equiv{1\over\sqrt{2}}\pmatrix{0&e^+\cr e^-&0}=\Phi+\Phi^\dagger$.
This object is a $U(1)$ invariant symmetric two-tensor.

The Polyakov path integral for pure gravity can be written in terms of
vielbeins as follows
\begin{equation}
\int {{\rm [space~of~invertible~vielbeins]}\over
{\rm [diffeomorphisms]\times [Lorentz]}}
\quad e^{-\mu\int e^+\wedge e^-}\quad,
\end{equation}
where $\mu$ is the cosmological constant.
\section{The $W-$string}
In this section, we will discuss the general case using $W_3$ to
provide more details. However, we shall try to preserve generality as
much as possible. Following Hitchin\cite{hitchin}, we consider the
vector bundle $V$ given by
\begin{equation}
V=S^{n-1}(K^{-{1\over2}}\oplus K^{1\over2})
=K^{-{{n-1}\over2}}\oplus K^{-{{n-3}\over2}}\oplus \cdots \oplus
K^{{{n-1}\over2}}
\end{equation}
and choose the Higgs field
\begin{equation}\label{ehiggsc}
\Phi=\pmatrix{   0   &   h    &   0  &\ldots& 0\cr
                {a_1 \over h}  & \ddots &\ddots&\ddots&\vdots \cr
              {a_{2}\over h^2}  & \ddots &\ddots&  h  & 0\cr
              \vdots & \ddots & {a_1\over h}  &  0  & h \cr
{a_{n-1}\over h^{(n-1)}}& \ldots & {a_2\over h^2}  & {a_1\over h} & 0}~dz\quad,
\end{equation}
where $a_i\in K^{i+1}$. $(V,\Phi)$ form a stable Higgs pair. Again
${\cal A}\equiv A + \Phi + \Phi^\dagger$ is a flat $PSL(n,\BC)$
connection provided the self-duality equations are satisfied. However,
the flat connection ${\cal A}$ has its holonomy contained in a split
real form of $PSL(n,\BC)$ which is isomorphic to $PSL(n,\BR)$.

Due to a theorem of Hitchin\cite{hitchin}, one of the connected
components of the space Hom$(\pi_1(\Sigma);PSL(n,\BR))/PSL(n,\BR)$
(which is the same as the space of solutions of the self-duality
equations for the Higgs pair $(V,\Phi)$ just described) has dimension
$(2g-2)~{\rm dim}~PSL(n,\BR) = (n^2-1)(2g-2)$ which is what one
expects as the dimension of the Teichmuller space of $WA_{n-1}$
gravity. Further, the Higgs field is parametrised by $a_i$ which
correspond to holomorphic quadratic, cubic, quartic, \ldots,
differentials\footnote{The zero modes of the $b$ anti-ghosts in the
conformal gauge are in one to one correspondence to quadratic and
higher differentials.}. We would like to identify this space with the
Teichmuller space of $WA_{n-1}$ gravity.

Following the bosonic string example in the previous section, we shall
now identify the Higgs field with {\it generalised vielbein} and the
gauge field $A$ with a {\it generalised spin-connection}. For example,
in the $WA_2$ string, the spin-connection is a $SO(3)$ connection and
the vielbein have five components. The $PSL(3,\BR)$ connection can be
parametrised as follows
\begin{equation}
{\cal A}^T =  \pmatrix{ \omega - {e^{+-}\over \sqrt{3}} &
(\omega^- + e^-) & \sqrt{2}e^{--}\cr
(\omega^+ - e^+) & {{2 e^{+-}} \over \sqrt{3}}&
(-\omega^- + e^-) \cr
\sqrt{2}e^{++}&(-\omega^+ - e^+) & - \omega - {e^{+-}\over
\sqrt{3}} }
\end{equation}
where $\omega, \omega^\pm$ form the $SO(3)$ connection and $e^A$ are
the generalised vielbein. We would like to make the following
observations. We shall provide details in a later publication\cite{later}.
\begin{enumerate}
\item[(i)] Here, $(e^\pm,$ $\omega)$ form a $PSL(2,\BR)$
conn\-ec\-tion which is em\-bed\-ded in $PSL(3,\BR)$.
\item[(ii)] $e^{++},e^{--},e^{+-}$ are the new vielbein. They have
been labelled by their $U(1)$ charges (w.r.t. $\omega$). Schoutens
{\it et al.} introduced W-vielbeins which appear to correspond to
$e^{++}$ and $e^{--}$ but not for $e^{+-}$ in their covariant
construction of an action for scalar fields coupled to $WA_2$
gravity\cite{schoutens}.
\item[(iii)] Relaxing the curvature zero condition on the $SO(3)$
spin-connection, we obtain a more general setting for $W_3$-geometry.
Just as the full general coordinate and Weyl transformations could be
recovered by relaxing the constant curvature condition, we can recover
the same in this case. W-Weyl transformations can be obtained by
suitably generalising a procedure due to Howe\cite{howe}.
\item[(iv)] In the $WA_2$ case, one can construct the ``metric''
$g_2$ from the quadratic $SO(3)$ invariant and a ``cubic tensor''
$g_3$ from the cubic $SO(3)$ invariant. This is done by introducing
$E$ which consists of only the vielbein terms in the $PSL(3,\BR)$
connection.  Then
$$
g_2 \sim tr(E^2)\quad,\quad g_3\sim tr(E^3)\quad.
$$
\item[(v)] The geometry is not Riemannian anymore in the sense that
the torsion constraints are not sufficient to determine the connection
in terms of the vielbein. For e.g., in the $WA_2$ case, there are 5
torsion constraints and the connection has 6 components. However, we
can always choose a gauge where we trade one of the gauge symmetries
to determine the connection in terms of the vielbein.
\item[(vi)] Conformal gauge corresponds to choosing
$$ e^+_z=e^-_{\bar z}=h\quad,\quad e^{+-}_z=e^{+-}_{\bar z}=0\quad, $$
which fixes the $SO(3)$ gauge freedom. Further, these gauge choices
are algebraic and hence their corresponding ghosts can be ignored
(since they would be non-interacting). Next, make the following
gauge choices
$$e^+_{\bar z}=e^-_z=e^{++}_{\bar z}=e^{--}_z=0\quad,$$
whose corresponding ghosts (anti-ghosts) are of spin $-1,-2$ $(2,3)$
as required for $WA_2$ gravity. Further, the residual transformations
which preserve this gauge choice correspond to holomorphic
(anti - holomorphic) transformations $\epsilon^+,\epsilon^{++}$
($\epsilon^-,\epsilon^{--}$).  The residual gravity degrees are freedom
are $h$, $e^{++}_z\equiv v^+$ and $e^{--}_{\bar z}\equiv v^-$.
\end{enumerate}
\section{$W$-diffeomorphisms}

In this section, we shall identify $PSL(n,{\BR})$ gauge
transformations with $W$ - diffeomorphisms provided a certain
constraint is satisfied.  Further, in contrast to earlier approaches,
we obtain $W$-diffeomorphisms in a purely geometric fashion without
invoking the presence of matter fields. In a different manner, the
work of Gerasimov {\it et al.}\cite{GLM} and subsequently that of
Bilal {\it et al.}\cite{BFK} discussed $W$-diffeomorphisms in the
conformal gauge, as deformations of certain flag manifolds associated
with jet bundles\footnote{These ideas have been further developed
subsequently by Zucchini\cite{zucchini}.}. In their construction, the
action of $WA_{(n-1)}$ diffeomorphisms on the vector space
$V=S^{n-1}(K^{-{1\over2}}\oplus K^{1\over2})$ was demonstrated.

\subsection{The $PSL(2,\BR)$ case}
As usual, the bosonic ($PSL(2,\BR)$) case shows us the way. Choose the
 Higgs field $\Phi$ as in eqn. (\ref{ehiggsb}). Consider the fields
 $(\tilde\psi_1,\tilde\psi_2)\in (K^{-{1\over2}},K^{1\over2})$ subject
 to the conditions
\begin{eqnarray}
 (d_\omega^{''} + \Phi^\dagger )\pmatrix{\tilde\psi_1\cr\tilde\psi_2}=0
\label{econda}\quad,\\
 (d_\omega^{'} + \Phi)\pmatrix{\tilde\psi_1\cr\tilde\psi_2}=0
\label{econdb}\quad.
\end{eqnarray}
Eqn. (\ref{econda}) implies that the fields $\psi_i$ are holomorphic
while the second eqn. (\ref{econdb}) is a constraint on $\psi_i$.
Interestingly, the self-duality equations (\ref{eselfdual}) and
(\ref{ehol}) are implied by the consistency of the two conditions we
have just imposed. Hence, self-duality equations correspond to the
integrability of equations (\ref{econda}) and (\ref{econdb}). The
holomorphicity condition (\ref{ehol}) on $\Phi$ implies that
\begin{equation}\label{econna}
\omega_z = -h^{-1}\partial_z~h\quad,\quad \omega_{\bar
z}=h^{-1}\partial_{\bar z}~h\quad,
\end{equation}
and that $a$ is holomorphic ($\partial_{\bar z}~a=0$).
We shall now rescale the fields $\psi_i$ as follows
\begin{equation}\label{escala}
\pmatrix{\tilde\psi_1\cr\tilde\psi_2} = \pmatrix{h^{1\over2}\psi_1\cr
h^{-{1\over2}}\psi_2} \quad.
\end{equation}
$(\psi_1,~\psi_2)$ correspond to {\it primary fields} in the CFT
sense, i.e., they have Einstein indices. This rescaling now enables us
to make contact with diffeomorphisms in CFT. Further, the Christoffel
connection are given by $\Gamma_{zz}^{~~z}=2\omega_z$ and
$\Gamma_{{\bar z}{\bar z}}^{~~{\bar z}}=2\omega_{\bar z}$. A simple
calculation shows that (\ref{econda}) and (\ref{econdb}) translate to
the following conditions
\begin{eqnarray}
(\partial_{\bar z} +{\mu}\partial_z - {1\over2}\partial_z
{\mu})\psi_1 =0\label{econdaa}\quad,\\
(\partial_z^2 -u)\psi_1 =0 \label{econdba}\quad,
\end{eqnarray}
where ${\mu}\equiv {{\bar a}\over{h^2}}$ is the Beltrami
differential and $u\equiv {1\over2}\partial_z\Gamma_{zz}^{~~z}
+{1\over4}(\Gamma_{zz}^{~~z})^2 -a$.  It is a simple exercise to check
that $u$ transforms like the Schwarzian. A similar observation was
made by Sonoda\cite{sonoda} who pointed out that such a term could be
added to the energy momentum tensor to make it transform like a
$(2,0)$ tensor. See also, \cite{eguchi}. Hence, $u$ behaves like a
projective connection. Eqn. (\ref{econdaa}) implies that $\psi_1$
transforms as
$$
\delta~\psi_1 = \xi^z\partial_z \psi_1 -{1\over2}(\partial_z \xi^z)\psi_1\quad,
$$
which is the standard transformation of a $(-{1\over2},0)$ tensor
in CFT.  This is somewhat similar to what has been done in
\cite{GLM,BFK}.  However, there are some differences. The flatness
condition that has been considered in \cite{GLM,BFK} is different from
the flatness condition implied by the self-duality equations. The
$PSL(2,\BR)$ gauge field is a completely geometric object with no
relation apriori to matter fields. However, a special combination of
the spin-connection transforms like the Schwarzian. This combination
can be related to the stress-tensor via Ward identities considered by
Verlinde\cite{verlinde,GLM,BFK}. It can also be seen that
$W-$transformations have a presentation here without directly
involving ``matter fields.''

The compatibility of conditions (\ref{econdaa}) and (\ref{econdba})
implies that
\begin{equation}\label{econsa}
[\partial_{\bar z} + {\mu}\partial_z + 2 (\partial_z {\mu})]~u =
{1\over2} \partial_z^3{\mu}\quad,
\end{equation}
which is equivalent to the standard OPE for the stress-tensor
(following a procedure outlined in \cite{GLM}.)
\begin{equation}\label{eopea}
T(z)~T(w) \sim {{c/2}\over {(z-w)^4}} + {{2T(w)}\over {(z-w)^2}} +
{{\partial_wT(w)}\over {(z-w)}} +\ldots\quad,
\end{equation}
provided we identify
$$ u \longrightarrow {6\over c} \langle T \rangle \quad.$$
Thus we have recovered the residual diffeomorphism in the conformal
gauge.

\subsection{The $W_3$ case}
We shall now repeat the exercise of the previous section for the
$WA_2$ case. Consider the multiplet
$(\tilde\psi_1,\tilde\psi_2,\tilde\psi_3)\in (K^{-1},K^0,K)$. The
Higgs field is given by (\ref{ehiggsc}) to be
$$
\Phi=\pmatrix{0&h&0\cr {a\over h}&0&h\cr {b\over {h^2}}&{a\over h} &0}dz
+ {\cal O}(a^2,ab,b^2)\quad.
$$
This corresponds to setting $v^\pm=0$ in the notation of section 3. We
shall not turn them on for simplicity.
The holomorphicity condition (\ref{ehol}) implies that
\begin{equation}
\omega^+_\mu = \omega^-_\mu =0 \quad,\quad \omega_z =-h^{-1}\partial_z
h\quad,\quad \omega_z =h^{-1}\partial_{\bar z} h\quad,
\end{equation}
and that $(a,b)$ are holomorphic ($\partial_{\bar z}a=\partial_{\bar
  z}b=0$).  The above solution is valid to ${\cal O}(a^2,ab,b^2)$.
Impose the following conditions on
$(\tilde\psi_1,\tilde\psi_2,\tilde\psi_3)$.
\begin{eqnarray}
 (d_\omega^{''} + \Phi^\dagger
)\pmatrix{\tilde\psi_1\cr\tilde\psi_2\cr\tilde\psi^3}=0
\label{econdc}\quad,\\
 (d_\omega^{'} + \Phi)\pmatrix{\tilde\psi_1\cr\tilde\psi_2\cr\tilde\psi_3}=0
\label{econdd}\quad.
\end{eqnarray}
Eqn. (\ref{econdc}) implies that the fields $\tilde\psi_i$ are holomorphic
while the second eqn. (\ref{econdd}) is a constraint on $\tilde\psi_i$. We
shall now rescale the fields $\tilde\psi_i$ as follows
\begin{equation}
\pmatrix{\tilde\psi_1\cr\tilde\psi_2\cr\tilde\psi^3}=
\pmatrix{h\psi_1\cr\psi_2\cr h^{-1}\psi^3}\quad.
\end{equation}
$(\psi_1,\psi_2,\psi_3)$ are the primary fields of CFT.
Conditions (\ref{econdc}) and (\ref{econdd}) translate to the
following conditions
\begin{eqnarray}
(\partial_{\bar z} -{ \mu}\partial_z + \partial_z{ \mu} + { \rho}
(\partial_z^2 -{2\over3}\tilde{u}_2) -{1\over2} \partial_z{
\rho}\partial_z +{1\over6}(\partial_z^2 { \rho}))\psi_1 =0
\label{econdca}\\
(\partial_z^3 - \tilde{u}_2\partial_z - (u_3 + {1\over2}
\partial_z\tilde{u}_2))\psi_1 =0\quad,\label{econdda}
\end{eqnarray}
where ${ \mu}\equiv {a\over h^2}$ and ${ \rho}\equiv {b\over
h^4}$ are the Beltrami differentials. Further,
$$
\tilde{u}_2 = 2\partial_z \Gamma_{zz}^{~~z} + (\Gamma_{zz}^{~~z})^2 + 2a
\quad,\quad u_3 = [\partial_z - 2\Gamma_{zz}^{~~z}](\omega_z^-h)-b\quad,
$$
where we have restored dependence on $\omega_z^-$ (even though it is
vanishing to ${\cal O}(a^2,ab,b^2)$ ). This is to show
that combinations of the connection behave like a generalised projective
connection in the sense of \cite{GLM,BFK}.

Equations (\ref{econdca}) and (\ref{econdda}) are identical to those
obtained in \cite[see eqn. (35)]{GLM}. As discussed earlier,
we can continue to use arguments identical to theirs and show that the
these two equations are equivalent to the following OPE's.
\begin{eqnarray}
T(z)~T(w) &=& {{c/2}\over {(z-w)^4}} + {{2T(w)}\over {(z-w)^2}} +
{{\partial_wT(w)}\over {(z-w)}} +\ldots\quad,\nonumber \\
T(z)~W(w) &=& {{3W(w)}\over {(z-w)^2}} +
{{\partial_w W(w)}\over {(z-w)}} +\ldots\quad,\nonumber \\
W(z)~W(w) &=& {{c/3}\over {(z-w)^6}} + \left({{2}\over {(z-w)^4}} +
{{\partial_w}\over {(z-w)^3}} +{3\over{10}}{\partial_w^2\over{(z-w)^2}}
\right. \nonumber \\
&&+\left. {1\over{15}}{\partial_w^3\over{(z-w)}}\right)T(w) 
+{{16}\over{5c}}\left({2\over{(z-w)^2}}+
{\partial_w\over{(z-w)}}\right)\Lambda(w)
\quad,\label{eopeb}
\end{eqnarray}
where $\Lambda=T^2$. The above OPE corresponds to the semi-classical limit
($c\rightarrow\infty$) of the OPE's of the $W_3$-algebra given
in \cite{zam} after we make the following identifications
$$
\tilde{u}_2 \longrightarrow {{24}\over c}\langle T \rangle\quad,\quad
u_3 \longrightarrow {{24}\over c}\langle W \rangle\quad.
$$

Equation (\ref{econdca}) implies the following transformation law for
a $(-1)-$ differential
\begin{equation}
  \delta~\psi_1 = \xi^z \partial_z \psi_1 - (\partial_z\xi^z)\psi_1 +
  \xi^{zz} (\partial_z^2 \psi_1) - {1\over2} (\partial_z\xi^{zz})
  (\partial_z\psi_1) + (({1\over6}\partial_z^2 - {2\over3}\tilde{u}_2)
  \xi^{zz})\psi_1 \quad,
\end{equation}
where $\xi^z$ and $\xi^{zz}$ (that parametrise $W$-diffeomorphisms)
are given by $\xi^z=\epsilon^+ h^{-1}$
and $\xi^{zz}=\epsilon^{++} h^{-2}$. $\epsilon^+$ and $\epsilon^{++}$
are $SL(3,\BR)$ gauge parameters.

We have demonstrated how $W-$diffeomorphisms are obtained in our
formulation. The interesting feature is that certain combinations of
the generalised connection play the role of projective connections.
This is not surprising since these connections are required to form
covariant derivatives under transformations which are much larger than
the set of projective transformations.

\section{Conclusion and Outlook}

In this letter, we have demonstrated that the Teichmuller spaces
${\cal T}_n(\Sigma)$ introduced by Hitchin as the space of solutions
of the self-duality equations are indeed the Teichmuller spaces for
$W$-gravity. Even though we have restricted ourselves to the case of
$WA_n$-gravity in this paper, this generalises to $W$-gravity related
to other semi-simple groups. By introducing generalised vielbeins and
connections, we are now able to give a gauge independent descriptions
of $W$-gravity. For the case of $W$-string theory, one can now write a
Polyakov path integral. This is done by considering the path integral
over the space of all generalised vielbein with the gauge symmetry
corresponding to generalised diffeomorphisms and generalised local
Lorentz transformations. The conformal and light-cone gauges can be
obtained by appropriately gauge fixing these symmetries. In order to
relate to the ``metric'' formulation\cite{hull,hullb}, we have suggested
that the symmetric tensors are given by higher order invariants.
$$
g_n\sim tr(E^n)\quad.
$$
We have explicitly shown how $W-$diffeomorphisms as discussed by
\cite{GLM,BFK} are reproduced in our formulation in the ``conformal
gauge.'' The interesting point is that special combinations of the
generalised connections played the role of projective connections.

We have seen that the generalised vielbeins and connections in
W-string theory can be represented by flat $PSL(n,\BR)$ connections
for the case of surfaces with genus $g>1$. What are the groups to be
chosen for the sphere and torus topology? We shall use the bosonic
case as well as the dimension of the group as a guide to figure out
the answer. In the bosonic case, the group is $SO(3)\sim SU(2)/Z_2$
for spherical topology and $ISO(2)\sim IU(1)$ for torus
topology\footnote{$IU(n)$ is the inhomogeneous group in $C^n$. It
consists of all the elements of $U(n)$ together with the group of
translations in $C^n$.}. This leads us to guess that the groups
$SU(n)$ for odd `$n$' and $SU(n)/Z_2$ for even `$n$' for spherical
topology and $IU(n-1)$ for torus topology. Both these groups have the
right dimension ($=n^2-1$). For example, $ISO(n)$ will not work in the
torus case except for $n=2$. Further, the torus case can be obtained
by contractions of either $SU(n)$ or $SL(n,\BR)$, just as in the
bosonic case\cite{gilmore}. It is probably essential to solve the
torus and sphere cases in detail for W-strings in this language in
order to understand the spectrum as well as the divergence structure
of these theories.

To make further progress towards a description of W-strings the most
important task seems to be understand the uniformisation associated
with the flat $PSL(3,\BR)$ connections. In the case of $PSL(2,\BR)$
this naturally leads us to the upper half-plane with the Poincare
metric and the Riemann surface is described by quotienting the upper
half-plane by the Fuchsian group. String matter fields can be
formulated as living on this space. The jet bundles of
$K^{-{1\over2}}$ are naturally associated with this
uniformisation\cite{selfdual}. In the case of $PSL(3,\BR)$ the natural
structure is the one associated with the 2-jet bundle of $K^{-1}$. The
uniformizing space in its most general form appears to be a space
which is a domain in $CP^2$. We might expect therefore in the
$PSL(3,\BR)$ case a domain in $CP^2$ which when quotiented by the
Fuchsian group (which is given by the embedding of the homotopy group
of the surface into a discrete sub-group of $PSL(3,\BR)$ ) describes
the Riemann surface plus the additional data encoded in the new
vielbeins and connections. This space would also naturally provide
$w-$ coordinates. This structure can also straightforwardly extended
to the general $W_n$ case\footnote{Gervais {\it et. al.} have
considered the possibility of surfaces embedded in $CP^n$ in the
context of W-geometry and Toda systems\cite{gervais,zucchini}. Also,
in a recent paper\cite{falqui}, the relationship of Toda systems to
Higgs bundles was demonstrated.}.

It is tempting however to try and relate the Hitchin construction for
$PSL(3,\BR)$ directly to Goldman's work on the convex real projective
($RP^2$) structures on Riemann surfaces\cite{rptwo}. In recent
work\cite{GC}, Choi and Goldman have shown that the Teichmuller space
of convex $RP^2$ structures is the same as the Teichmuller component
of the moduli space of Higgs bundles for $PSL(3,\BR)$.  Goldman also
provides Fenchel-Nielsen co-ordinates for his Teichmuller space by
considering convex domains in $RP^2$ quotiented by a Fuchsian subgroup
of $PSL(3,\BR)$. In contrast to the $PSL(2,\BR)$ case, there are two
lengths (and corresponding twists) associated to every $PSL(3,\BR)$
non-conjugate element of the Fuchsian group. $(6g-6)$ of these lengths
( and their twists) together with two other parameters associated with
every pant in the corresponding pants decomposition provide the
$(16g-16)$ real co-ordinates of the Teichmuller space of convex $RP^2$
structures. The gluing prescription is also generalised to the $RP^2$
case, involving however only the lengths and twists.  Some progress
has also been made towards using these co-ordinates to write down a
Weil-Petersson like measure on the Teichmuller spacer, which contains
within it the usual Weil-Petersson measure on ${\cal T}_2$.

Even though the Higgs bundle approach appears to lead to an
uniformisation by a domain in $CP^2$, the associated real holonomy
should lead to a construction identical to that of Goldman. The
challenge here is to formulate the matter part of W-string theory on
this uniformizing space and provide the kind of explicit description
available in standard string theory.  In analogy with the bosonic
string\cite{DP,gava}, it would be interesting to study the structure
of W-string theory at the boundaries of this Teichmuller space.

\noindent {\bf Acknowledgements:} We would like to thank M. Blau, S. Das, S. F.
Hassan, K. Joshi, P. Majumdar, D. S. Nagaraj, K.S. Narain, S. Nag,
B. Rai, P.  Sastry, A. Sen and S. Wadia for useful
discussions. S. G. would like to thank the organisers of the
International Colloquium on Modern Quantum Field Theory II at Tata
Institute, Bombay (January 1994) and the Spring Workshop on Strings,
Quantum gravity and Gauge theories at ICTP, Trieste (April 1994) for
an opportunity to present some of these results.


\begin{thebibliography}{99}
\bibitem{zam} A. B. Zamolodchikov, Teor. Mat. Fiz. {\bf 65} (1985)
1205. \\
V. A. Fateev and A. B. Zamolodchikov, Nuc. Phys. {\bf B280} (1987) 644.
\bibitem{hitchin} N. J. Hitchin, Topology {\bf 31} (1992) 451-487.
\bibitem{selfdual} N. J. Hitchin, Proc. London Math. Soc. {\bf 55} (1987)
59-126.
\bibitem{dbg} J. de Boer and J. Goeree, Nuc. Phys. {\bf B401} (1993)
369.
\bibitem{hull} C. M. Hull, Commun. Math. Phys. {\bf 156} (1993) 245;
Nuc. Phys.{\bf B 413} (1994) 296.
\bibitem{bouwknegt} F. A. Bais, P. Bouwknegt, K. Schoutens and M.
Surridge, Nuc. Phys. {\bf B304} (1988) 348.
\bibitem{falqui} E.Aldrovandi and G.Falqui,
``Geometry of Higgs and Toda Fields on Riemann Surfaces,''
Preprint hep-th/9312093.
\bibitem{GLM} A. Gerasimov, A. Levin and A. Marshakov, Nuc. Phys. {\bf B360}
(1991) 537-558.
\bibitem{BFK} A. Bilal, V. Fock and I. Kogan, Nuc. Phys. {\bf B359} (1991)
635-672.
\bibitem{others} de Boer and J. Goeree, Nuc. Phys. {\bf B381} (1992) 329-359;
Phys. Lett. {\bf B274} (1992) 289-297. \\
K. Yoshida, Int. J. of Mod. Phys. {\bf A7} (1992) 4353-4376.
\bibitem{corlette} K. Corlette, J. Diff. Geom. {\bf 28} (1988) 361-382.
\bibitem{donald} S. K. Donaldson, Proc. London Math. Soc. {\bf 55} (1987)
127-131.
\bibitem{tftreview} D. Birmingham, M. Blau, M. Rakowski and G. Thompson,
Physics Reports {\bf 209} (1991) 129-340.
\bibitem{topgrav} D. Montano and J. Sonnenschein, Nuc. Phys. {\bf B324}
(1989) 348.\\
E. Verlinde and H. Verlinde,  Nuc. Phys. {B348} (1991) 435.
\bibitem{verlinde} H. Verlinde, Nuc. Phys. {\bf 337} (1990) 652.
\bibitem{schoutens} K. Schoutens, A. Sevrin and van Nieuwenhuizen,
Phys. Lett. {\bf B243} (1990) 245-249; Nuc.  Phys. {\bf B349} (1991) 791-814.
\bibitem{howe} P. S. Howe,  J. Phys. {\bf A12} (1979) 393.
\bibitem{zucchini} R. Zucchini, Class. Quant. Grav. {\bf 10} (1993) 253-278.
\bibitem{later} S. Govindarajan and T. Jayaraman (in preparation).
\bibitem{sonoda} H. Sonoda, Nuc. Phys. {\bf B281} (1987) 546.
\bibitem{eguchi} T. Eguchi and H. Ooguri, Nuc. Phys. {\bf B282} (1987) 308-328.
\bibitem{gilmore} See chapter 10 in R. Gilmore, {\it Lie Groups, Lie
Algebras and some of their applications}, John Wiley \& Sons, New
York.
\bibitem{hullb} For instance, see C. M. Hull, ``Classical and Quantum
W-gravity,'' in the proceedings of the summer school in High Energy Physics
and Cosmology, 1992 (World Scientific) and references therein.
\bibitem{gervais} J-L. Gervais and Y. Matsuo, Commun. Math. Phys. {\bf 152}
(1993) 317-368.
\bibitem{rptwo} W. M. Goldman, J. Diff. Geom. {\bf 31} (1990) 791-845.
\bibitem{GC} S. Choi and W. M. Goldman,  Proc. of the AMS {\bf 118}
(1993) 657.
\bibitem{DP} E. D'Hoker and D. H. Phong,  Rev. Mod. Phys. {\bf 60} (1988) 917.
\bibitem{gava} E. Gava, R. Iengo, T. Jayaraman and R. Ramachandran,
Phys. Lett. {\bf B168} (1986) 207.
\end{thebibliography}
\end{document}